\documentclass[twoside,english,12pt]{JAC2003}

\usepackage{amssymb}
\usepackage{graphicx}
\usepackage{multicol}

\begin{document}

\title{Secondary electron yield measurements from thin surface coatings for NLC
electron cloud reduction}
\author{F. Le Pimpec, F. King, R.E. Kirby, M. Pivi \\
SLAC, 2575 Sand Hill Road, Menlo Park, CA 94025 }
\maketitle

\thispagestyle{headings} \markright{SLAC-PUB-10438}

\begin{abstract}

In the beam pipe of the positron damping ring of the Next Linear
Collider, electrons will be created by beam interaction with the
surrounding vacuum chamber wall and give rise to an electron
cloud. Several solutions are possible for avoiding the electron
cloud, without changing the bunch structure or the diameter of the
vacuum chamber. Some of the currently available solutions for
preventing this spurious electron load include reducing residual
gas ionization by the beam, minimizing beam photon-induced
electron production, and lowering the secondary electron yield
(SEY) of the chamber wall. We will report on recent SEY
measurements performed at SLAC on TiN coatings and TiZrV
non-evaporable getter thin films.

\end{abstract}

\footnotetext{Work supported by the U.S. Department of Energy,
Contract DE-AC03-76SF00515}

%%%%%%%%%%%%%%%%%%%%%%%%%%%%%%%%%%%%%%%%%%%%%%%%%%%%%%%%%%%%%%%%%%%
\vspace{-0.2cm}

\section{Introduction}

\vspace{-0.2cm}

Beam-induced multipacting, which is driven by the electric field
of successive positively charged bunches, arises from a resonant
motion of electrons that were initially generated by photon, gas
ionization or by secondary emission from the vacuum wall. These
electrons move resonantly along the surface of the vacuum chamber,
occasionally getting "kicked" by the circulating beam to the
opposite wall. The electron cloud effect (ECE), due to this
multipacting, has been observed or is expected at many storage
rings. The space charge of the cloud, if sufficient, can lead to a
loss of the beam or, at least, to a drastic reduction in
luminosity. In order to minimize the electron cloud problem which
might arise in the NLC, we are looking to solutions involving
surface coating of the secondary electron emitting vacuum wall.

\section{Experiment Description and Methodology}

\begin{figure}[tbph]
\begin{center}
\vspace{-0.3cm}
\includegraphics[width=0.5\textwidth,clip=]{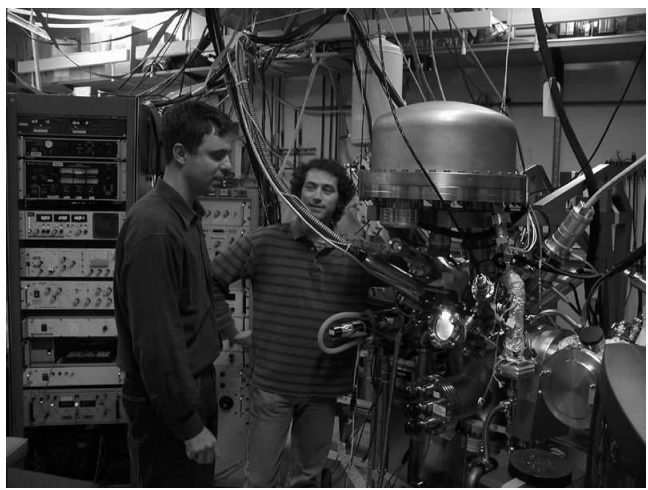}
\end{center}
\vspace{-0.6cm}
\caption{Experimental system used for SEY measurements and surface analysis}
\vspace{-0.2cm}
\label{SEYkirbysetup}
\end{figure}

The system and methodology used to measure the secondary electron
yield has been described thoroughly in reference \cite{lepimpec:LCC128}. The
system is composed of two coupled stainless steel UHV chambers
where the pressure is in the low 10$^{-10}$ Torr scale in the
measurement chamber and high 10$^{-9}$ Torr scale in the "load
lock" chamber, Fig.\ref{SEYkirbysetup}. Samples individually
screwed to a carrier plate, are loaded first onto an aluminium
transfer plate in the load lock chamber, evacuated to the low
10$^{-8}$ Torr scale, and then transferred to the measurement
chamber.

\begin{figure}[tbph]
\begin{center}
\includegraphics[width=0.4\textwidth,clip=]{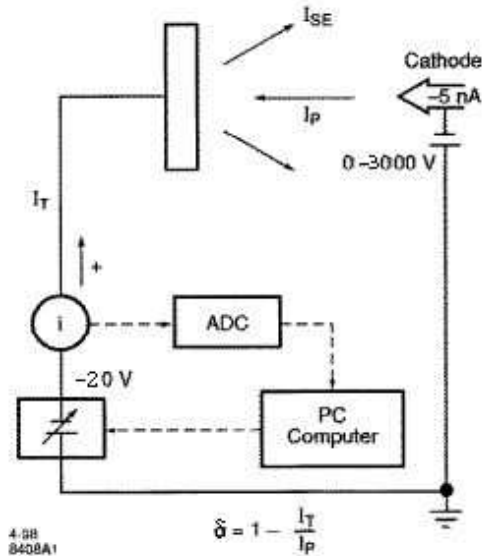}
\end{center}
\caption{Electronic circuitry used to measure the secondary
emission yield}
\label{figelctrqcircuit}
\end{figure}

One sample at a time is measured in the measurement chamber, by
placing it on a dedicated support, Fig.\ref{figelctrqcircuit}.
Sample temperature can be monitored by the use of two
thermocouples, and chemistry by x-ray photoelectron spectroscopy
(XPS). In the case of a non-evaporable getter (NEG) sample,
thermal activation of the layer is provided by electron
bombardment to the back of the sample.

SEY measurements were made with a Keithley 6487, a high resolution
picoameter with internal $\pm$505~V supply and IEEE-488 interface.
The 6487 has several filter modes which were turned off for our
measurements. The integration time for each current reading is set
to 167~microsec, which is the minimum  value for the instrument. The
current for each primary energy step was sampled one hundred times; the mean and standard deviation
were returned from the picoameter to the computer.

\vspace{-0.2cm}

Calculation of the SEY ($\delta$) is done via the equation in
Fig.\ref{figelctrqcircuit}

%equation~\ref{equSEY}.
%It is used because it contains parameters measured directly in the
%experiment.

%\begin{equation}
%\delta = \frac{Number\ of\ electrons\ leaving\ the\
%surface}{Number\ of\ incident\ electrons}
%\label{equdefinition}
%\end{equation}

%\begin{equation}
%\delta = 1 -\frac{I_T}{I_P}
%\label{equSEY}
%\end{equation}

where I$_P$ is the primary lectron gun current impinging on the  sample and I$_T$
is the total current measured on the sample ($I_T = I_P + I_S$).
I$_S$ is the secondary electron current leaving the sample.

It is important to not look at the SEY at low primary energy and
try to conclude something about elastic reflectivity. Data below
20~eV comes from a band structure and are a combination of
diffraction from the crystalline structure and energy absorption
by the material \cite{kirby:1985}. Surface effects such as
roughness, angles of incidence of the primary electron and
chemistry on the surface influence the SEY of a material. More
details on the methodology can be found in reference \cite{lepimpec:LCC128}

\vspace{-0.2cm}

\section{Results \& comments}
\vspace{-0.2cm}

\begin{figure}[tbph]
\centering
\includegraphics[width=0.5\textwidth,clip=]{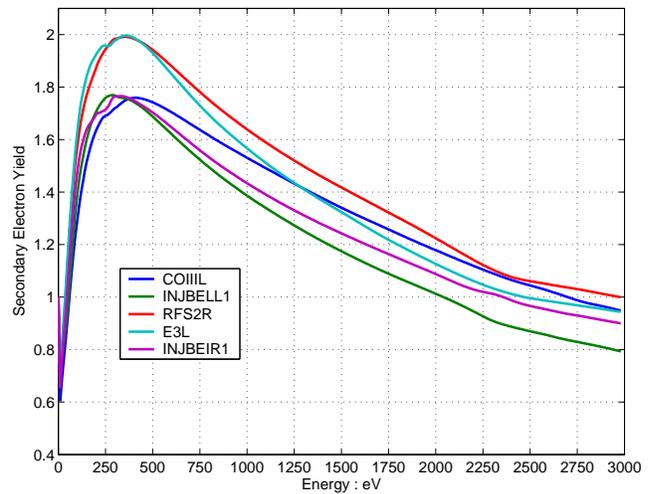}
%\setcaptionwidth{6cm}
\caption{SEY of five TiN/SS samples, as received}
\label{figTiNSS0-3kev}
\end{figure}

 As also shown in reference \cite{lepimpec:LCC128} for
TiN/Al; the SEY of TiN/SS (TiN coated on stainless steel) has a
spread, see Fig.\ref{figTiNSS0-3kev}. The process used to coat
them is described in reference \cite{he:PAC01}. The spread in the
$\delta_{max}$ can be hypothesized as depending on contamination,
roughness, and nitrogen pressure \cite{Hseuh:ecloud04}.
Contamination and stoichiometry determination, of the samples,
were obtained by XPS, cf Table.\ref{XPStablle}.

\vspace{-0.1cm}

\begin{table}[tbph]
\begin{tabular}{|c|c|c|c|}
  \hline
  Sample & Ti At\% & N At\% & Contamination \\
  \hline
  INJBEIR1 & 14 & 15.4 & - \\
  INJBELL1 & 20 & 23.75 & - \\
  RFS2R & 14.5 & 14.9 & Sodium \\
  E3L & 13.4 & 13.06 & - \\
  CO111L & 16 & 14.6 & Sodium \\
  \hline
\end{tabular}
\caption{XPS survey of TiN/SS sample} \label{XPStablle}
\end{table}

\vspace{-0.1cm}

\begin{figure}[htbp]
\centering
\includegraphics[width=0.5\textwidth,clip=]{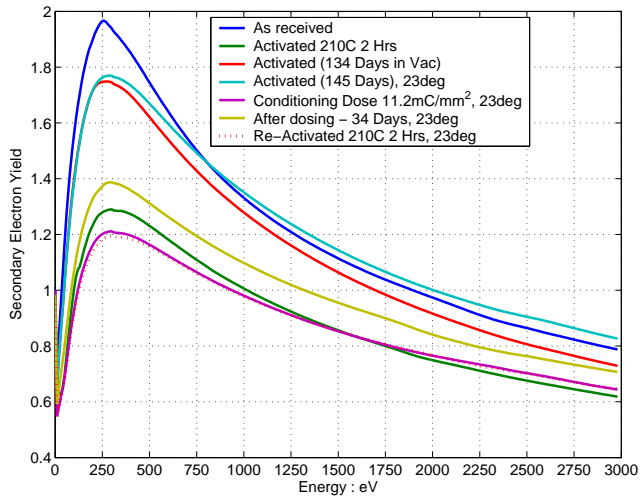}
\caption{SEY of TiZrV after different processes}
\label{figTiZrV0-3kev}
\end{figure}

\begin{figure}[htbp]
\centering
\includegraphics[width=0.5\textwidth,clip=]{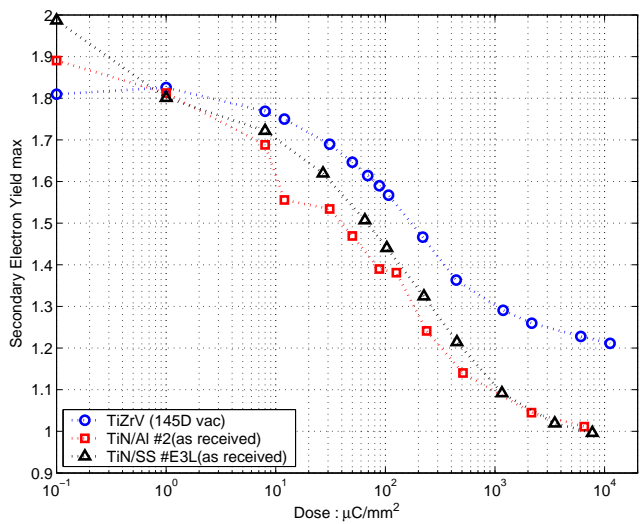}
\caption{SEY max during electron conditioning of TiZrV, TiN/Al, and TiN/SS}
\label{figSEYmaxdose}
\end{figure}

\begin{figure}[htbp]
\centering
\includegraphics[width=0.5\textwidth,clip=]{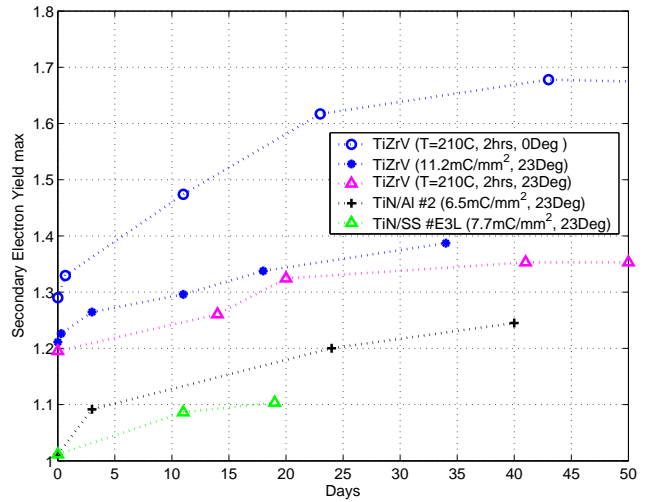}
\caption{SEY max during recontamination in a vacuum of few 10$^{-10}$~Torr}
\label{figSEYmaxdays}
\end{figure}

Results, for a series of processes, from a $Ti_{30}Zr_{18}V_{52}$,
at\%, NEG coated on SS substrates are shown in
Fig.\ref{figTiZrV0-3kev}. Initially, the sample was measured "as
received", then after a first activation at 210$^\circ$C for 2
hours. The sample is then left in the measurement chamber for 145
days at a pressure below 10$^{-9}$~Torr,~N$_2$ equivalent. The
system was then exposed frequently to the unbaked vacuum of the
load lock chamber of a few 10$^{-9}$~Torr. The next step was
bombardment of the sample by electrons of kinetic energy 130~eV.
Results of this electron surface conditioning are shown in
Fig.\ref{figSEYmaxdose}. This conditioning effect is also observed
for the TiN/Al and TiN/SS samples, Fig.\ref{figSEYmaxdose}. The
NEG sample is then left in vacuum for 34 days before being
thermally reactivated at 210$^\circ$C for 2 hours. Effects of the
recontamination by this residual vacuum below 10$^{-9}$~Torr on
the $\delta_{max}$, for the TiN and NEG samples after these
different processes, are shown in Fig.\ref{figSEYmaxdays}.

\vspace{-0.3cm}

\vspace{-0.6cm}

\vspace{-0.45cm}

\vspace{-0.45cm}

\section{Conclusion}

\vspace{-0.2cm}

We have presented a brief report on the status of SEY experiments
carried out at SLAC. In the case of the NEG getter coating, the
influence of the activation and recontamination on its pumping
action were investigated. The maximum SEY $\delta$ increased from
$\sim$1.2 to $\sim$1.4 after forty days of exposure to a vacuum of
$\sim$5.10$^{-10}$~Torr. The second set of data after activation
agree with CERN measurements\cite{Scheurlein:2002}. Gas-saturated
and conditioned NEG seems to not have a $\delta_{max}$ above 1.4.
Conditioning the NEG with a 130~eV electron beam leads to a
$\delta_{max}$ of 1.3, after a dose of 1~mC/mm$^2$. The influence
of electron conditioning has been shown for the TiN on SS or Al
substrate. Values of $\delta_{max}$, reached at a dose of
1~mC/mm$^2$, are 1.1 for both samples. Recontamination does not
degrade the SEY dramatically, Fig.\ref{figSEYmaxdays}.

\vspace{-0.3cm}

\section{Acknowledgments}

\vspace{-0.2cm}

%We would like to thank P.~He and H.C.~Hseuh, BNL for providing
%the TiN samples and C.~Benvenuti at CERN for the TiZrV sample.
%Most valuable was the work of G. Collet and E. Garwin, SLAC, for
%converting and baking the XPS system for use on SEY measurements.

We would like to thank P.~He and H.C.~Hseuh at BNL for providing
the TiN samples and C.~Benvenuti at CERN for the TiZrV sample.
Thanks to G. Collet and E. Garwin, SLAC, for adding the SEY
capability into the XPS system.

%%%%%%%%%%%%%%%%%%%%%%%%%%%% References %%%%%%%%%%%%%%%%%%%%%%%%%

%\bibliographystyle{unsrt}
%\bibliography{flp_slac}

\end{document}